\documentclass[aps,prl]{revtex4}
\usepackage{epsfig}
\usepackage{bm}
\usepackage{graphicx}
\usepackage{amssymb}
\usepackage{amsmath}
\usepackage{color}
\usepackage{slashed}
\newcommand{\ra}{\rangle}

\newcommand{\ben}{\begin{displaymath}}
\newcommand{\een}{\end{displaymath}}
\newcommand{\be}{\begin{equation}}
\newcommand{\ee}{\end{equation}}
\newcommand{\bea}{\begin{eqnarray}}
\newcommand{\eea}{\end{eqnarray}}

\newcommand{\eq}[1]{Eq.~(\ref{#1})}

\def\r{\rho}

\def\b{\beta}
\def\s{\sigma}\def\O{\Omega}
\def\La{\Lambda}

 \newcommand{\csb}{{\rm \; charge\; symmetry\; breaking\;}}

\begin{document} 
\title{\bf \hskip10cm \\
{Ernest Henley's Isospin and the Ensuing Progress } %{Application to nuclear charge radii} 
%\\ --or-- \\
%{the neutron skin thickness} \\
%--or-- \\
%{Ca isotopes}
}
%\newcommand*{\UW}{Department of Physics, University of Washington, Seattle, WA 98195}
%\newcommand*{\UWindex}{1}
%\affiliation{\UW} 

\author{G.A. Miller}
 \affiliation{Department of Physics,
University of Washington, Seattle, WA 98195-1560}
\date{\today}

\begin{abstract}
Ernest Henley's contributions to understanding isospin symmetry and the experimental and theoretical progress that followed  are reviewed. Many experimentalists and theorists worked to bring Ernest's early vision of the subject to fruition. This progress and wide-ranging impact is described.  \end{abstract}
 
 \maketitle 
\noindent

\begin{figure}[htb]
\centering
\includegraphics[height=3.5in,angle=-2.25]{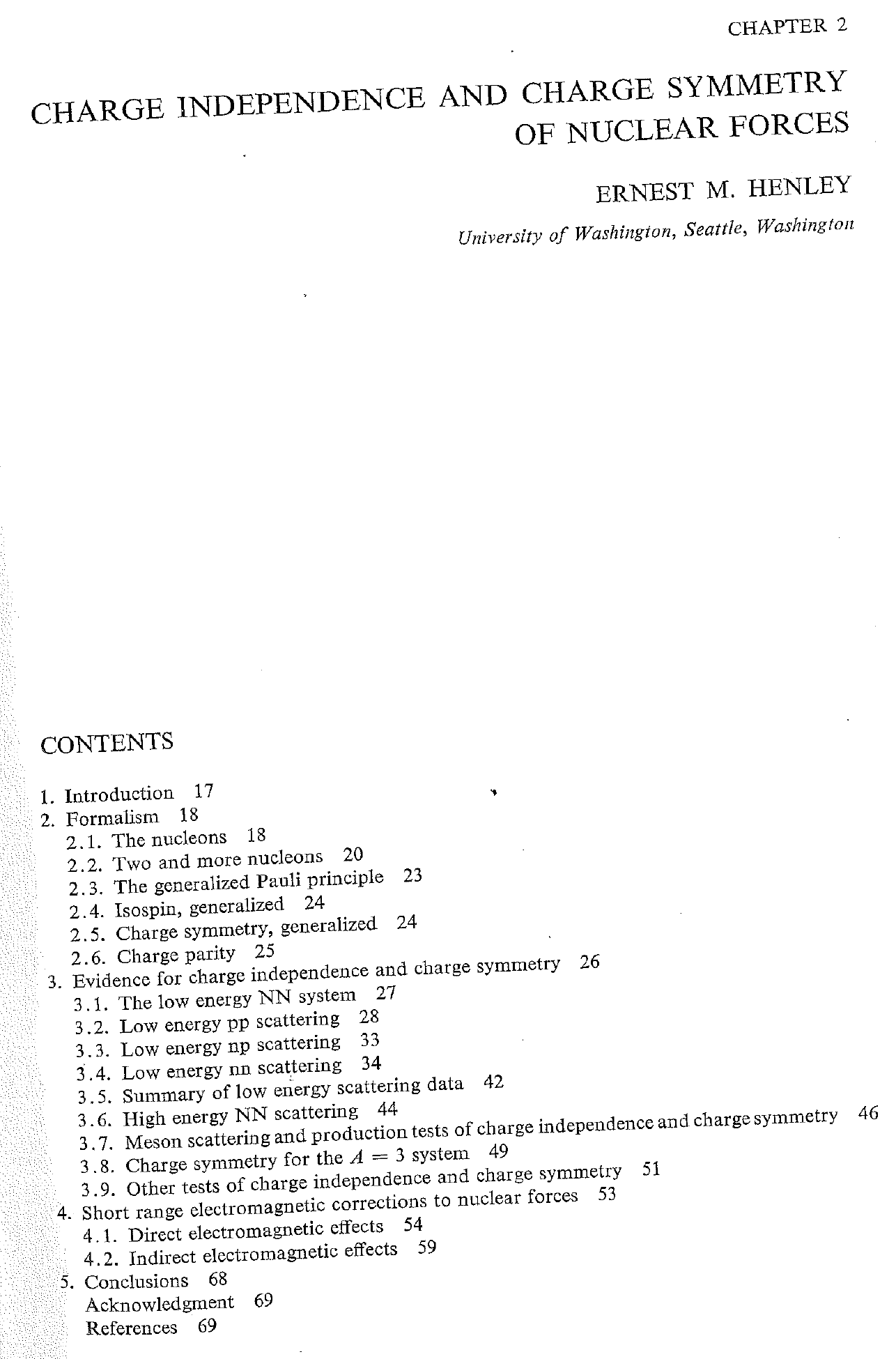}
\caption{First page of Ernest Henley's 1970 article}
\label{EMHarticle}
\end{figure}

I was very happy to contribute to this volume intended to honor the contributions of Ernest Henley, a man who had a  very major influence on my career. My first encounter with Ernest occurred about two years before I met him.  I was a graduate student at MIT, studying  isospin violating hadronic corrections to the widths of isobaric analog states, working with Arthur Kerman.  Some time around 1970, John Negele told me that there was a new preprint on isospin violations from Henley~\cite{Henley:1970}  in the Physics Library.
The article was much more advanced than any on that topic that I  had ever seen. I thought: What is a man like that doing in Seattle? I was soon to find out!
 
I think that it is fair to say that Ernest established the field of isospin as a fundamental symmetry. The first page of the article,  shown in Fig.~\ref{EMHarticle}, displays   the scope of his work. Note that it was Chapter 2, the first real chapter after the Introduction, in a 750 page book of 14 chapters. This paper established isospin as an approximate symmetry, collected the evidence for this statement, and originated various terms. Indeed Ernest coined the term ``approximate symmetry" to denote a symmetry that was slightly broken by small effects.

Ernest organized the nuclear force. It is a definite fact that the neutron-proton (np) and nn nuclear forces are different, but these forces were called charge independent! Ernest defined charge independence as a symmetry in which the $pp$, $np$ and $nn$ forces are equal to each other in the same space-spin state.
For nuclear forces, charge symmetry was defined as the identity of $nn$ and $pp$ forces.  
Ernest  generalized these statements to all hadrons.  Electromagnetic forces break  these symmetries, as expected. Ernest also  catalogued the non-electromagnetic forces.

 One detail that proved very important involves the difference between isospin invariance (charge independence) and charge symmetry. Isospin invariance means that the physics is independent of any rotation in isospin space. For this to be true one needs $[H,T_i]=0,$ where $T_i$ is a component of the isospin operator.  
 Charge symmetry is invariance under a particular rotation: by 180$^\circ$ about the $y-$axis, if the $z$-axis is associated with charge:
 \bea P_{\rm cs}\equiv e^{i \pi T_2}.\label{Pcs}\eea
 Charge symmetry does not imply isospin invariance. The breaking of charge symmetry does imply the breaking of charge independence.
 
 Ernest defined future directions for the field. Regarding theory, he wrote ``Unless an understanding of the hadronic NN force is possible, it will remain extremely difficult 
 to make convincing theoretical calculations of non-electromagnetic hadron-hadron interactions."   For experiment he advocated further work to accurately establish the $nn$ scattering length. In particular, he focussed on the reaction
 $\pi^-+D\rightarrow nn+\gamma$. The theory was carried out brilliantly by Gibbs, Gibson \& Stephenson~\cite{Gibbs:1975ew}. This was the first paper I saw that presented an error analysis of the theory. Later measurements of the reaction~\cite{Gabioud:1979uz,Chen:2008zzj} led to establishing the fact that the $nn$ interaction is more attractive than the $np$ one in the $^1S_0$ state.
\section{Next Steps}
\begin{figure}[htb]
\centering
\includegraphics[height=3.49in]{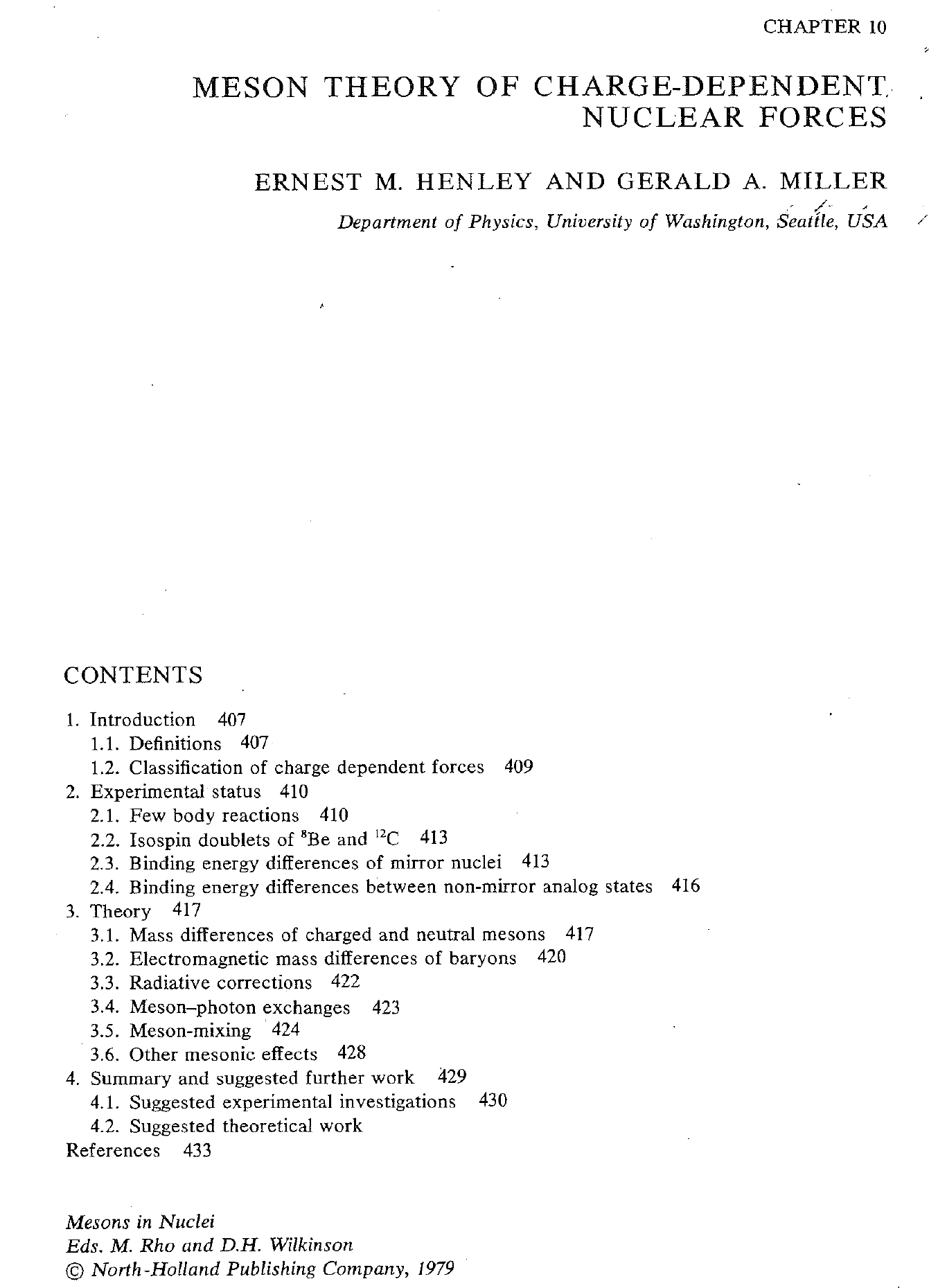}
\caption{First page of our 1977 article~\cite{Henley:1977qc}}
\label{HM79}
\end{figure}

I arrived in Seattle in Sept. 1975. In the following January Ernest asked me to join in  writing a review~\cite{Henley:1977qc}.  He said that charge independence breaking was well-established, therefore the focus should be on  charge symmetry breaking, CSB. My memory is that Ernest had almost the complete outline written  at the time he asked me to help. I supplied some technical details. The wide scope of the article is apparent from the first page, Fig.~\ref{HM79}. Along with  the review,  many papers on this subject ultimately  resulted. The earliest were Refs.~\cite{Cheung:1978it,Alberg:1978wc,Cheung:1979ma}.
 I also eventually wrote further reviews \cite{Miller:1990iz,Miller:1994zh,Miller:2006tv}.

Ernest's summary of the theory was interesting. He wrote    that CSB was difficult to calculate, but at least it could stimulate experiment.  For experiment he/we  had three suggestions. One could test the center of mass polarization  relation $P_n(\theta)=P_p(\pi-\theta)$. Charge symmetry would be broken if the cross section for $dd\rightarrow ^4{\rm He}+\pi^0 $ reaction were measured to be non-vanishing. Another proposed test was the angular asymmetry about 90$^\circ$ in the center of mass of the reaction $np\to d\pi^0$.

Ernest was not content to merely publish  papers. He actively worked with the committees that chose experiments at the TRIUMF and IUCF laboratories. I later worked with experimenters at those labs on related theoretical issues.
The experimental progress on these reactions is discussed below, but first it worthwhile to provide some detail regarding isospin breaking in the nucleon-nucleon system.

\section{Henley-Miller nucleon-nucleon force classification (1977)}\label{sec:hm77}
We review the CSB and CIB terminology of nucleon-nucleon forces~\cite{Henley:1977qc}.

Class (I): Forces which are isospin independent that commute with all components of the isospin operator. Such forces, $V_I$ have an isoscalar  form,
\begin{equation}
V_I= a+b \vec{\tau}(i)\cdot\vec{\tau}(j),
\label{eq1}
\end{equation}
where $a$ and $b$ are Hermitian  isospin independent operators and $i\ne j$.

Class (II): Forces which maintain charge symmetry, but break charge independence. These can be written in
isostensor form,
\begin{equation}
V_{II}=c(\tau_3(i)\tau_3(j)- \vec{\tau}(i)\cdot\vec{\tau}(j),
\label{eq2}
\end{equation}
The Coulomb interaction leads to a Class II force as do the effects of the pion mass difference in pion exchange forces.
Effects of charge-dependent coupling constants may also lead to such a Class II force. 

Class (III): Forces which break both charge independence and charge symmetry, but which are symmetric under the interchange $i\leftrightarrow j$ in isospin space,
\begin{equation}
V_{III}=d(\tau_3(i)+\tau_3(j)).
\label{eq3}
\end{equation}
A  Class III force differentiates between $nn$ and $pp$ systems, but does not cause isospin mixing in the two-nucleon system because
\begin{equation}
[V_{III},T^2]=0.
\label{eq4}
\end{equation}
The effects of $\rho^0-\omega$ mixing yields such a force, as does the Coulomb interaction.

Class (IV) Class IV forces break charge symmetry and therefore charge dependence; they cause isospin
mixing. These forces take the form
\begin{equation}
V_{IV}=e(\vec{\sigma(i)}-\vec{\sigma(j))}\cdot \vec{L}(\tau_3(i)-\tau_3(j))+f(\vec{\sigma}(i)\times\vec{\sigma}(j))\cdot \vec{L}(\tau_3(i)\times\tau_3(j)),
\label{eq5}
\end{equation}
where $\vec{L}$ is the two-nucleon orbital angular momentum  $e$ and $f$ are Hermitian isospin-independent operators.
Such forces give CSB spin-orbit effects that account for the np analyzing power differences~\cite{Abegg:1986ei,Abegg:1988kx,Abegg:1995qi,Abegg:1998sg,Knutson:1991vj,Vigdor:1992pb} and contribute to nuclear isospin mixing~\cite{Wiringa:2013fia}.

\section{Search for Class IV forces}
Charge symmetry leads to the equality of the differential   differential cross sections for polarized neutrons scattering from unpolarized protons and vice versa. As a result $A_n(\theta)= A_p(\theta)$ where 
$A_{n,p}$  denotes the analyzing power and where the subscript represents the polarized nucleon. (Time reversal invariance dictates that $A(\theta)=P(\theta).$)  A non-vanishing asymmetry difference is directly proportional to the isotopic spin singlet-triplet, spin singlet-triplet mixing amplitude and therefore direct evidence of a Class IV force.  The first measurement of charge symmetry breaking in $np$ elastic scattering  (at an incident neutron energy of 477 MeV) was performed at
 TRIUMF~\cite{Abegg:1986ei,Abegg:1988kx} by  measuring the difference  $ A_n -A_p$, at the zero-crossing angle of the average analyzing power. A  non-zero result, which was largely explained by the effect of the neutron-proton mass difference on the one-pion exchange potential~\cite{Miller:1986mk,Williams:1987yn}, was observed.  This discovery was followed by
 confirmation by an accurate IUCF experiment~\cite{Knutson:1991vj,Vigdor:1992pb}. Later TRIUMF work at 347 MeV~\cite{Abegg:1995qi,Abegg:1998sg} also found a significant effect.
In general, the measured analyzing power differences of the IUCF and TRIUMF experiments were well reproduced by theoretical predictions based on the paradigm at the time: meson exchange potential models, which indirectly incorporate quark level effects. The calculations include contributions from one photon exchange (the magnetic moment of the neutron interacting with the current of the proton), from the neutron-proton mass difference affecting charged single $\pi,\r$   and $\omega$ exchange, and from   isospin mixing $\r^0-\omega$  meson exchange. We note that all of the model parameters related to the strong interaction were constrained fits to phase shifts, so that the calculations were testing only  the inputs related to charge symmetry breaking. 
The review~\cite{Miller:1994zh}  discusses this topic in more detail.

%\cite{Abegg:1986ei,Abegg:1988kx,Abegg:1995qi,Abegg:1998sg,Knutson:1991vj,Vigdor:1992pb}
\section{The reaction $dd\rightarrow \alpha \pi^0$}
Ernest Henley's words ``if isospin is conserved this reaction is forbidden, however contrary to all claims made in the literature, this reaction only tests charge parity" are very interesting. Ernest was willing to call out  his colleagues in general, but was not particularly disposed to publicly naming names. Charge parity was Ernest's  phrase  for charge symmetry.

A successful measurement of the reaction was made at IUCF~\cite{Stephenson:2003dv} after several incomplete attempts at other laboratories.  The authors  reported the first observation of the charge symmetry breaking  $dd\rightarrow \alpha \pi^0$ reaction near threshold. by  measuring  total cross sections for neutral pion production of $12.7 \pm  2.2$ pb at 228.5 MeV and $15.1 \pm  3.1 $ pb at 231.8 MeV. These cross sections arise fundamentally from the down-up quark mass difference and quark electromagnetic effects that contribute in part through meson mixing.
These values of the cross section were qualitatively reproduced by theoretical calculations~\cite{Gardestig:2004hs,Nogga:2006cp,Lahde:2007nu,Fonseca:2009qs}.  Note the small value of the measured cross sections. This arises because the cross section is amazingly the  {\it  square}  of a charge symmetry breaking amplitude. All other measurements of charge symmetry breaking are essentially first-order effects.

The IUCF program was followed further observations at higher energies performed at the WASA-at-COSY facility~\cite{Adlarson:2014yla,Adlarson:2017sid}.
The review~\cite{Miller:2006tv} discusses this topic in   detail.

\section{The reaction $np\to d\pi^0$}
\begin{figure}[htb]
\centering
\includegraphics[height=2.25in]{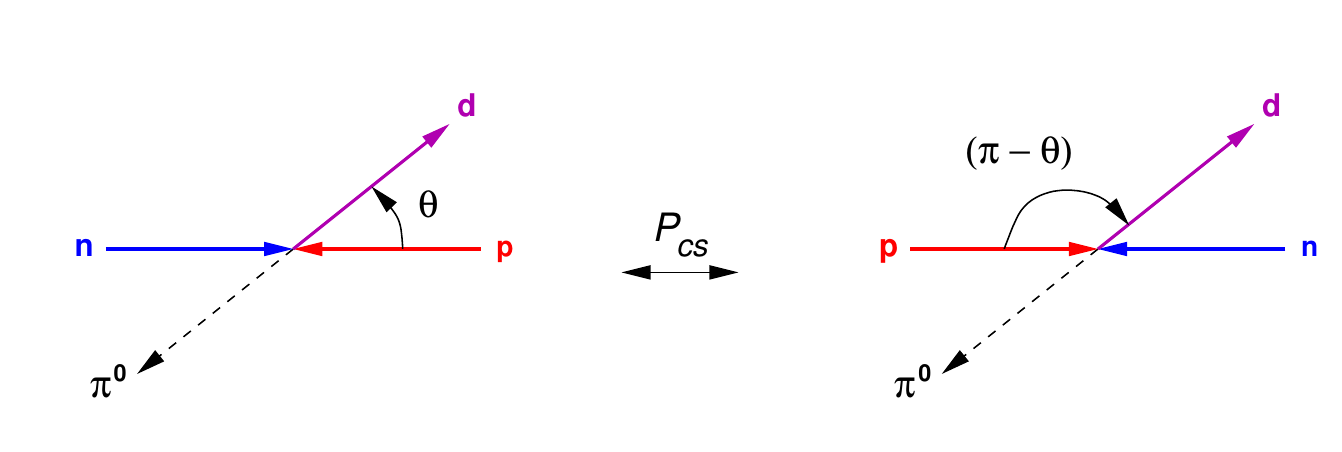}
\caption{Charge symmetric relations for the reaction $np\to d\pi^0$ in the center of mass frame. The rotation $P_{cs}$ (\eq{Pcs})  rotates the neutron into a proton and vice versa}
\label{npdpi0}
\end{figure}

An  asymmetry about 90$^\circ$ in the center-of mass (cm) frame in the reaction $np\to d\pi^0$ can only be caused by charge symmetry breaking. This is shown in Fig.~\ref{npdpi0}.  The size of the charge symmetry breaking can be inferred from the cm $np\to d\pi^0$ forward-backward asymmetry, $A_{\rm fb}$,  defined as
\bea
%A_{\rm fb}(\theta)={\s(\theta)-\s(\pi-\theta)\over \s(\theta)+\s(\pi-\theta)}, 
A_{\rm fb}\equiv {\int_0^{\pi/2}d\O\,\left(\s(\theta)-\s(\pi-\theta)\right)\over \int_0^{\pi}d\O\,\s(\theta)}
,\eea where $\theta$ is the cm angle between the incident neutron beam and the scattered deuteron.
A successful measurement was made ~\cite{Opper:2003sb} in 2003. The review~\cite{Miller:2006tv} discusses this topic in more detail.
\section{Isospin in QCD}
QCD was yet to be  discovered at the time of Henley's 1970 review.   
The notion that quantum chromodynamics (QCD), with its quarks and gluons, is the fundamental theory of the strong interactions  generated many new perspectives for studying hadronic systems and interactions. In QCD the interactions of up   $(u$) and down $(d)$ quarks are identical, if the $uÑd$ quark mass difference is neglected. This identity  of interactions is termed charge symmetry. The definition of the charge symmetry operation $P_{\rm cs}$  is given in \eq{Pcs}.  In QCD  $u$ and $d$ quarks carry the isospin quantum numbers and the charge symmetry
operator changes $d$ into $u$ quarks, and $u$ into $d$ quarks~\cite{Miller:1990iz}
\bea
P_{\rm cs}|d\ra=|u\ra,\,P_{\rm cs}|u\ra=-|d\ra.
\eea
In this language, the quark mass difference plus electromagnetic effects is responsible for charge symmetry breaking. 
It is also true that in this language, the quark mass difference plus electromagnetic effects is responsible for {\it all} of isospin breaking, so that the distinction between charge symmetry breaking and isospin breaking becomes less pronounced. Nevertheless, the distinction between charge symmetry breaking and isospin violation   remains very important at the hadronic level. For example, the mass difference between the neutron and the proton is an example of charge symmetry breaking (CSB), but the $\pi^\pm-\pi^0$ mass difference is not.

In QCD the only sources of CSB are
electromagnetic interactions and the mass difference between up and down quarks, so CSB studies are necessarily concerned with the origins and consequences of the small quark mass difference. Therefore, the relationship between the fundamental 
quark-gluon dynamics and  observable hadronic effects is a natural goal of the CSB studies discussed in previous sections.
The QCD basis of CSB is discussed in detail in the review~\cite{Miller:1990iz}. The consequences of the up-down quark mass difference, as elaborated via effective field theory are discussed in the review~\cite{Miller:2006tv}. Some of that material is recounted in the next Section.

\section{Isospin in Effective Field Theory}
Effective field theory, EFT,  is the modern technique to evaluate the consequences of QCD for hadronic and nuclear physics.
This is possible because the most general Lagrangian 
which respects unitarity, has correct properties 
under cluster decomposition and the same symmetries as QCD, should be equivalent
to  QCD
\cite{Weinberg:1978kz}.
 Thus one may use chiral hadronic Lagrangians, 
 to make QCD predictions at relatively low energies
\cite{vanKolck:1999mw,Bedaque:2002mn,Hanhart:2003pg}.
Predictive power is retained when  at low energies
 an expansion in momentum, formulated using power counting arguments, 
can be shown to
converge.  

The principal aim of EFT  as applied to nuclear physics is to handle
the short distance  (or  high momentum)  physics, the region that most concerned Ernest Henley.  One often resorts to using models to 
 deal with
short-distance  ($\approx 0.7$ fm)  interactions interactions between two or more nucleons.
  EFT offers a systematic procedure  of
 treating 
 ultraviolet-divergent integrals that appear in evaluating loop diagrams by using
cutoffs at a scale $\La$ and counter terms with magnitude expressed in terms 
of low energy constants LECs that depend implicitly on $\La$.
 If the Lagrangian and renormalization procedure is 
 correctly constructed,   computed observables are independent of  $\La$. 
  In making calculations 
one expands  low-energy observables,  that depend upon a momentum scale $Q$,
 in
powers of the hopefully small parameter $Q/\Lambda \ll 1$ and 
argues that only a finite number of LECs
contribute at a
given order in the $Q/\Lambda$ expansion. Thus one hopes to obtain
a Lagrangian  with a  small
number of coefficients that can be determined from 
experimental data,  fundamental theory,  or models. The resulting Lagrangian can then be
used
 to predict other observables.  

 %ain
%how one can do QCD without quarks and gluons.
To  understand  charge symmetry breaking in nuclear systems, the 
Lagrangian should be 
expressed in terms of  pions, nucleons, $\Delta$'s, and photons.
Unitarity and correct cluster decomposition must be retained in any effective theory,
but the essential feature of QCD is its 
 approximate chiral symmetry (for small quark masses), which
imposes important restrictions on the way pions interact.
 The quark mass difference  breaks isospin explicitly. 
In  low-energy EFT, the effect of quark-mass terms
can be reproduced if
 all terms that break chiral symmetry in the same way are included.
These interactions can involve  the pionic fields   without derivatives,
but are always accompanied by powers of $m_u+m_d$
or $m_u-m_d$.
 
%\cite{bkm,vanKolck_review}.

 \subsection {EFT and pion production}

The production of a single pion  in a nucleon--nucleon collision  has
a rather special role because of its relation with 
chiral symmetry.  Furthermore, it is a source of $\pi^0$
that can access the potentially large effects of charge symmetry breaking~\cite{Weinberg:1994tu} (see below).
One might therefore hope that effective field theories will give insights
and allow the determination of $m_d-m_u$.

 Chiral symmetry
  provides
 the conditions for the construction of an effective field theory, in
that it forces  
the mass of the pion $m_\pi$, as the Goldstone boson of the chiral
symmetry breaking, to be low, and  the interactions to be weak.  
In the chiral limit ($m_\pi=0$) the pion must be  be free of interactions for vanishing
momenta.  

\subsubsection{EFT and CSB}
\label{sub:csbeft}
%This part will discuss the role of EFT in CSB -- the Weinberg-van Kolck 
%treatments 
%of effective $\pi N$ scattering operators, the definitions of  criteria 
%for convergence, etc.
%The emphasis will be on use of EFT in pion production 
%in both the strong interaction and CSB.

The mass difference between $u$ and $d$ quarks breaks isospin symmetry
explicitly. 
Indication from the meson masses is that
the ratio $\varepsilon \equiv (m_u-m_d)/(m_u+m_d) \sim 1/3$. Naively,
this suggests that isospin might not be a much better symmetry than
the rest of the chiral group. On the other hand, a quick 
look at hadron masses and much experimental evidence show that isospin is typically broken
only at the few percent level.

That  isospin is such a good symmetry can be understood
from  the 
pattern of chiral-symmetry breaking incorporated in 
the chiral Lagrangian \cite{invk}.
While explicit chiral-symmetry-breaking effects are present
already at the lowest order in $Q/\La$  through the pion mass term,
operators generated by the quark mass difference appear only at the next order
  through a term that
contributes to the nucleon mass splitting 
and,
due to chiral symmetry, to certain pion-nucleon interactions. 
As a consequence, in most quantities
isospin breaking competes with isospin-conserving
operators of lower order,
and its relative size is not $\varepsilon$
but $\varepsilon (Q/M_{QCD})^n$, where $n$ is a positive integer. 
In other words, isospin is an accidental symmetry \cite{invk}:
a symmetry of the lowest order EFT which is not a symmetry
of the underlying theory.

The only known exception to this rule is in the isoscalar
$t$ channel in $\pi N$ scattering at threshold. In this case,
 there is no contribution from the leading-order  Lagrangian, and
both the isospin-conserving and -violating amplitudes start
at the same order. The isospin-violating
piece comes from the pion-nucleon interactions
linked to the nucleon mass splitting. As described in previous sections, the effects have been 
  seen experimentally, but many possible terms involving different manifestations of the up-down quark mass difference contribute to the scattering amplitude.
%finish

Note that the classes I-IV of  a previous section %ct.~\ref{sec:hm77}
 have been shown to correspond to the increasing orders of chiral perturbation theory~\cite{invk,vanKolck:1996rm}. Class I is lowest order, Class II  is of the next order, and so on.

\section{CSB nucleon electromagnetic form factors and Parity Violating (PV) Electron Scattering}

The basic idea is that elastic  PV electron-proton scattering is sensitive to nucleon strangeness content~\cite{Armstrong:2012bi}, and also  to the value of the weak-mixing angle~\cite{Androic:2013rhu,Androic:2018kni}.  So far, a convincing signal for strangeness in the nucleon has not  been seen~\cite{Armstrong:2012bi}. 

The relevance of charge symmetry or its breaking to PV electron scattering on the proton  arises from the need to relate the amplitude for $Z$-boson absorption on the proton to measured proton and neutron electromagnetic form factors. This
can be done only if  charge symmetry holds to an extent that its breaking  is much smaller than any expected contribution from strangeness.

 The breaking of charge symmetry brings in a correction that cannot be obtained directly from experimental observations
~\cite{Dmitrasinovic:1995jt,Miller:1997ya,Kubis:2006cy}.  The key question is whether the uncertainty in obtaining the correction is large compared to current and projected experimental uncertainties. Experimentalists  stated that charge symmetry  limits the ability to push further on the strange form factors because results obtained with improved precision would be hard to interpret cleanly in terms of strangeness or CSB.

I addressed~\cite{Miller:1997ya} the question of whether or not CSB really limits the ability to push further. That paper found  that the CSB corrections are less than 1\% of the size of the electromagnetic form factors $G_E,G_M$.
When re-expressed in terms of absolute values of charge symmetry breaking form factors, the results were very small
of order 2$\times10^{-3}$. This is small enough to ignore, despite the amazingly high  accuracy of current experiments.  

However, I had ignored the effect of charge symmetry breaking arising from the influence of the neutron-proton mass difference on the pion cloud of the nucleon. This effect was included by Kubis \& Lewis~\cite{Kubis:2006cy}. These effects   are not small because of a log divergence in the loop integrals.
In their resonance-saturation procedure the pion graph is cut off at the mass of the rho meson, and rho-omega mixing graphs provide a finite counter term. This is larger than the pion diagrams.  The result, the charge symmetry breaking magnetic form factor ranges between 0.01 and 0.04, or about 10 times larger than my result. There is also a large uncertainty in the results due to lack of knowledge of the $\omega$ nucleon strong tensor coupling.  

  Kubis \& Lewis~\cite{Kubis:2006cy} take the strong coupling constants from dispersion analyses of electromagnetic form factors based on vector meson dominance. Such fits are well known to be flexible~\cite{Matevosyan:2005ia,Matevosyan:2005bp}. The strong coupling constants for
 omega nucleon coupling are about  seven times larger than used in NN scattering. So there is a conflict. 
 
 How can  we tell which method (or if either method) is correct? One answer is that the effects of rho-omega mixing in nucleon-nucleon scattering is constrained. It is known to give a medium range class III CSB  potential  that can account for the scattering length difference between $nn$ and $pp$ systems~\cite{McNamee:1974vb,Miller:1990iz}, and a class IV CSB potential that plays an important role in understanding 
 CSB in np scattering. The class III potential may account for the missing binding energy difference between $^3He$ and $^3H$~\cite{Coon:1987kt} and also the Nolen-Schiffer anomaly~\cite{Blunden:1987wr}, see the review~\cite{Miller:1994zh}. There are other possible sources of CSB that influence these energy differences. All effects have the same sign because the are all driven by the mass difference between up and down quarks.  The use of the KL coupling constants gives  potentials that are rather different, and much larger, than the one ~\cite{Coon:1987kt} needed phenomenologically.

 We~\cite{Wagman:2014nfa}  made later  calculations of the CSB form factors using relativistic chiral perturbation theory. The use of relativistic chiral perturbation theory leads to finite and convergent results. The CSB effects were found to be    an order of magnitude smaller than current experimental bounds on proton strangeness. 
 
 More generally, I wish to address a bias. I did a quark {\it model } calculation. Kubis \& Lewis did a chiral perturbation {\it theory} calculation. One usually thinks that a theory is better than a model. However, if an unconstrained counter term is needed to evaluate the theory, then a model that  is constrained by experimental data is better than  a theory. 

\section{Other subjects}

I want to briefly mention three other topics that are  relevant to more modern topics. 

\subsection{Super allowed Fermi $\b$ decay}

In my opinion,  the computations of the rate for nuclear  super allowed beta decay, used  to test the unitarity of the CKM matrix, could be improved. In particular, we~\cite{Miller:2008my,Miller:2009cg} used an exact formalism to  show that certain radial excitations that are often neglected are significant. My impression is that it is difficult to include such effects in ab initio calculations. Nevertheless, given the importance of the subject, I hope that  someone will look further at this.

\subsection{The NUTEV anomaly}
  The NuTeV group\cite{Zeller:2001hh}
 has measured charged 
and neutral current reactions for deep inelastic scattering of
neutrinos by iron targets. 
Ratios of these cross sections  can be used to determine
the Weinberg angle, provided that the target is isoscalar,
the nuclear strangeness content can be
ignored, charge symmetry holds, and a variety of nuclear effects on
 deep inelastic scattering can be neglected. In this case the
Paschos-Wolfenstein  \cite{Pas73} relation holds and $\sin^2\theta_W$ can be extracted from ratios of neutral current to charged current  
 total 
cross sections, averaged with the proton and neutron. 
 This simple relation follows from equating the
$u,d$ contributions in the proton scattering with those of the
$d,u$ contributions to the neutron scattering, and then 
taking
the difference
between the neutrino and anti-neutrino cross 
sections. This  isolates the axial-vector interference term 
  proportional to 
${1\over 2}-\sin^2{\theta_W}$ for the neutral current. 
The NuTeV value for $\sin^2 \theta_W $ is three standard 
deviations larger than the value measured in other electroweak processes.
However  $\csb$
causes a change in the Paschos-Wolfenstein  relation~\cite{Sather:1991je,Rodionov:1994cg,Londergan:2003pq}, and this correction has been shown to be
proportional to a ratio of
 quark momenta that is independent of $Q^2$ and 
essentially model-independent~\cite{Londergan:2003ij}. 
 The size of the correction
is about 50\% of the discrepancy 
 with other determinations of the Weinberg angle \cite{Londergan:2003ij}. There are other nuclear and hadronic effects that complicate the interpretation of this reaction.  See {\it e.g.} \cite{Miller:2002xh}

\section{$m_d-m_u>0$ causes all strong-interaction  charge symmetry breaking}

There is a very definite striking  pattern caused by the positive value of  $m_d-m_u>0$. The neutron mass is greater than the proton mass, in contrast with the expectations based on electromagnetic effects. But this is not all, as detailed in 
Table 1.1 of Ref.~\cite{Miller:1990iz}. The mass difference between the members of  dozen hadronic isospin doublets may be understood by counting the number of up and down quarks. This pattern occurs for mesons and baryons. Moreover, this mass difference is responsible for the mixing of mesons.  Electromagnetic effects also violate charge symmetry, but in most cases these are smaller than the strong-interaction effects of $m_d-m_u>0$.  

A recent paper~\cite{Gibbs:2017non}  uses a model in which ``the Coulomb interaction is the source of all isospin breaking". The basic idea of this work is that at short center-of -mass separations   between two nucleons ``the six quarks intermixed give more possibilities for correlations among them leading to additional Coulomb-energy."
This may be true, but the neglect of the effects of the quark mass difference is a very significant omission. Such give rise to significant effects in nucleon-nucleon scattering and to mass differences between mirror nuclei~\cite{Koch:1985yi}. In particular that reference found that ``the computed mass difference between six quark clusters formed by two neutrons and two protons is less than that (2.6 MeV) of free nucleons in both the non-relativistic and MIT bag quark models. This is because quark-quark Coulomb and hyperfine interactions have an increased magnitude in the six-quark system. This makes a neutron-rich nucleus slightly more bound than its proton-rich mirror. The magnitude of the effect is of about the right size to account for the missing Coulomb energy problem (Nolen-Schiffer anomaly) for nuclei with A=3, 13, 17, 29, 33, and 41." 
The six-quark clusters have very different structure than two-nucleons, being mainly hidden color configurations.
That paper also  shows that the Coulomb effects give attraction, but the effects of the light quark mass difference on the kinetic energy and the hyperfine interaction are of the same order of magnitude. The numerical values depend strongly on what we called the six-quark bag probability. This is the probability that two-nucleons make a close enough encounter so that it is correct that the quark degrees of freedom can be used. We had some a sensible  prescription~\cite{Henley:1983ag}, but unfortunately there is not yet any convincing evidence that this probability is non-vanishing.  This affirms what everyone knows: the quark presence in nuclei is difficult to establish. It is hard to separate quark effects from those obtained from other mechanisms.

The work of  Ref.~\cite{Gibbs:2017non}  initiates a model of nucleon-nucleon scattering that is consistent with scattering lengths. This is then used to compute the $A=3$ binding energy difference, using a wave function that is expressed in terms of nucleon coordinates However, it is well known that using {\it any} potential  that accounts for the scattering length difference will account for the $A=3$  binding energy difference. Thus 
the computed agreement for $A=3$ provided in this reference provides no (nada, nil, rien de tout) evidence for the particular Coulomb  mechanism used to create the potential. Furthermore,
  the claim of~\cite{Gibbs:2017non}  that the Coulomb interaction is exclusively  responsible for the nucleon-nucleon scattering length differences  is incomplete. There are other mechanisms that are non-zero.   Most probably, their underlying non-relativistic quark  model is not sufficiently accurate for the problem at hand.  The claims in the abstract are not supported by sufficient evidence.
 
 In my opinion, modern tools such as lattice gauge theory calculations will eventually settle the issue. It has  already been calculated~\cite{Borsanyi:2014jba} that the influence of  up-down quark mass difference on the $n-p$ mass difference is about 2.5 times larger in magnitude  than that of electromagnetic effects.

\section{Concluding remarks}

Each of the previous sections ends with some sort of a conclusion, so I'd like to wrap up by making further remarks about Ernest Henley.
Ernest's isospin vision, laid out in his reviews, was carried out by a host of experimentalists and theorists.  This program  was carried out with great success and much was learned. It is perhaps  more important to recognize that 
Ernest always acted with grace in difficult situations. His actions provided a model of the correct way to behave. I  have tried to emulate his behavior. I am very grateful for having known him and will
miss him as long as it is possible.
 
\section*{Acknowledgements}

   This work was supported by the U. S. Department of Energy Office of Science, Office of Nuclear Physics under Award Number  DE-FG02-97ER-41014.

%\bibliography{references,eep,emc}

 \end{document}